# A NOVEL SOLUTION TO THE DYNAMIC ROUTING AND WAVELENGTH ASSIGNMENT PROBLEM IN TRANSPARENT OPTICAL NETWORKS

Urmila Bhanja[1], Sudipta Mahapatra[2], Rajarshi Roy[3]

[1]Research Scholar, Indian Institute of Technology, Kharagpur, India.
`urmilabhanja@gmail.com`
[2]Associate Professor, Indian Institute of Technology, Kharagpur, India.
`sudipta@ece.iitkgp.ernet.in`
[3]Assistant Professor, Indian Institute of Technology, Kharagpur, India.
*royr@ece.iitkgp.ernet.in*

*ABSTRACT:*

*We present an evolutionary programming algorithm for solving the dynamic routing and wavelength assignment (DRWA) problem in optical wavelength-division multiplexing (WDM) networks under wavelength continuity constraint. We assume an ideal physical channel and therefore neglect the blocking of connection requests due to the physical impairments. The problem formulation includes suitable constraints that enable the algorithm to balance the load among the individuals and thus results in a lower blocking probability and lower mean execution time than the existing bio-inspired algorithms available in the literature for the DRWA problems. Three types of wavelength assignment techniques, such as First fit, Random, and Round Robin wavelength assignment techniques have been investigated here. The ability to guarantee both low blocking probability without any wavelength converters and small delay makes the improved algorithm very attractive for current optical switching networks.*

## 1. INTRODUCTION:

Among various issues in optical networks, the routing and wavelength assignment (RWA) problems are quite important for efficient network operations. The RWA problem in networks employing wavelength division multiplexing (WDM) determines the wavelength channel for each request. We focus on the dynamic RWA problem, which is more challenging than its static counterpart. In the dynamic RWA problem, since connection requests arrive randomly, it is more difficult to establish the light paths. Generally, dynamic RWA algorithms aim to minimize the total blocking probability in the entire network. Both static and dynamic RWA problems in any WDM network are NP-complete and many heuristic approaches have been proposed in the literature for these problems [1]. These problems have also been solved using linear programming (LP) or integer LP (ILP) methods [2, 3]. Surveys of routing and wavelength assignment algorithms are found in Zang et al. [4]. The authors have shown in their paper that to make the RWA problem computationally simple, it is usually divided into two sub-problems that are solved separately: the routing problem and the wavelength assignment problem. All most all the papers dealing with the DRWA problem in the literature follow adaptive routing [5] where the routing is dependent on the current network state information.

As far as the wavelength assignment part is concerned, among the available algorithms the random and first-fit (FF) techniques are the most practical as they are simple to implement [4]. Another wavelength assignment strategy that may be adopted is the Round-Robin strategy [6]. All the above wavelength assignment techniques do not require global knowledge of the network. Many other heuristic approaches for wavelength assignment are also found in the literature, such as the least used, most used, least loaded, and MAX-SUM [4].

Recent literature shows that there are many bio-inspired algorithms proposed to solve dynamic RWA problems in WDM networks. In [7, 8] the authors formulated both the static RWA and dynamic RWA problems as multi-objective optimization problems and solved them using genetic algorithm. Their approach can solve the static RWA problem very well; but, it is not suitable for a network with highly dynamic traffic. Bisbal et al. [9] proposed a novel GA-based algorithm for the DRWA problem. The algorithm proposed by them gives low blocking probability and a low computation time. The drawback of their method is that they have considered only path length in their fitness function. Le et al. [10] have proposed an improved GA to solve a dynamic RWA problem. For better load balancing in the network, they have designed a new fitness function. They have achieved a lower blocking probability than that of the genetic algorithm proposed by Bisbal et al. [9]. Vinh Trong Le et al. [11] have proposed another algorithm based on combination of mobile agents and genetic algorithm. They have proposed a fitness function that simultaneously takes into account the path length, number of free wavelengths, and wavelength conversion capability in route evaluation. Recently, some papers have adapted a combination of Ant Colony Optimization with Particle Swarm Optimization to solve dynamic RWA problem under the wavelength continuity constraint [12].

All of the above mentioned papers solve the DRWA problem for Poisson connection request arrival process and exponential holding time and consider minimizing the blocking probability only. Bisbal et al. [9] have also tried to minimize average execution time along with the average blocking probability. In the design of the DRWA problem not only is the blocking performance important, but also the execution time should be low, to be suitable for real time applications. Hence, there is a need for developing a fast algorithm for the DRWA problem to be used in online applications. To the best of our knowledge, till date there is no EP (evolutionary programming) based solution to the DRWA problem that considers the set up time of lightpaths while deciding the best possible path. Set up time of lightpaths is the time taken by the initialization and the mutation process of the algorithm to find a feasible chromosome or the time taken to find a feasible lightpath. Our work is to develop a novel solution to the DRWA problem based on an evolutionary programming algorithm. We also compare the dynamic performance of the network for three different wavelength assignment strategies using a Poisson process to model the arrival of user initiated ftp session connection requests, or TCP session connection requests, in a wide area network. We have observed the network performance for both exponential holding time and Pareto holding time for the DRWA problem. Exponential holding time models the call holding time in circuit switched telecommunication network and Pareto holding time models the session holding time on a web site.

The main contributions of our work are the following:

- As in the proposed algorithm the initial population consists of only a single chromosome, population initialization takes much less time compared to a standard genetic algorithm.

- The computational complexity is greatly reduced as mutation alone does the network state exploitation and exploration as suggested by Fogel [13].

- The fitness function formulation integrates information regarding the path length, the set up time of lightpaths, and the free wavelengths availability in the route; thus, it

International journal of Computer Networks & Communications (IJCNC), Vol.2, No.2, March 2010

provides a good load balancing and low blocking probability, while incurring a low execution time in finding a feasible route.

The paper is organized as follows. The problem model and formulation is presented in Section 2. The proposed algorithm as well as the implementation details of the algorithm for the DRWA problem is explained in Section 3 and 4 respectively. Section 5 gives the experimental results along with the related discussions. Finally, Section 6 concludes the paper

## 2. PROBLEM MODEL AND FORMULATION:

In the DRWA problem the lightpath requests are assumed to arrive at the network dynamically according to a Poisson process at an average arrival rate of $\lambda$. The connection request for a lightpath is specified by three parameters: the source node, S, the destination node, D, and the holding time for that particular request, $T_h$. The source-destination pair for each lightpath request is randomly generated according to a Uniform distribution and the holding times for the lightpath requests are assumed to be either exponentially distributed or Pareto distributed with some fixed mean 't_hold'

### 2.1. Network and Routing Models:

The 14 node NSF network is considered to illustrate the proposed algorithm is modeled as a graph G(V, E), where V is the set of nodes, representing routers or optical cross connect switches (OXCs), and E is the set of bidirectional fiber links representing physical connectivity between the nodes. Eight wavelengths are assumed to be provided per fiber.

In our routing model we make use of a variable $I_{ij}^{lp}$, which tells us whether link *(i, j)* is used by the lightpath *lp*. When the link *(i, j)* is used by the lightpath *lp*, $I_{ij}^{lp}$=1; otherwise, $I_{ij}^{lp}$=0. We will consider this to be a positive variable when the lightpath link leaves the node, and negative in the opposite case. A lightpath from the source *S* to the destination *D* is represented as *path(lp)* and is a collection of all the links belonging to the lightpath from *S* to *D*. There is a variable $C_{i,j}$ associated with each link *(i,j)*$\in E$, representing the overall cost of transmitting a packet over that link. LP is the set of all the light paths. The details of the proposed EP approach and wavelength assignment are given in the following sections.

### 2.2. Problem Formulation:

As mentioned before the proposed EP based algorithm represents the path finding process as a constrained optimization problem. The adopted objective function and the constraints are specified as follows:

**Objective function:** Here the objective is to dynamically find a cost optimal route that maintains wavelength continuity over all the involved links. The objective is achieved by maximizing the fitness function defined for the *x-th* chromosome by:

$$f_x = \frac{W_x}{\sum_{j=1}^{k_x-1} C_{gx(j),gx(j+1)}} + \frac{W_x}{\sum_{(i,j)\in E} H_{ij}} + \frac{W_x}{T_x} \qquad (1)$$

where $W_x$ is the free wavelength factor explained in Section 2.3, $k_x$ is the length of the *x-th* chromosome, $g_{x(j)}$ represents the gene of the *j-th* locus of the *x-th* chromosome and $g_{x(j+1)}$ represents the gene of the *(j+1)-th* locus of the *x-th* chromosome. $C_{gx(j), gx(j+1)}$ represents the cost of transmitting a packet between the *jth* and *(j+1)th* locus of the *x-th* chromosome $H_{ij}$ is the number of hops between the *i-th* and *j-th* nodes, and $T_x$ represents the setup time of a request for the *xth* chromosome. In this work, an optimal path or route means the route that minimizes the routing cost, hop count and set up time of a path while adhering to the flow conservation constraints. Care also has been taken to check that the paths or routes generated are must be without any loops.

### 2.3. Wavelength Assignment Model:

We use three different wavelength assignment approaches First-Fit, Round Robin, and Random techniques and compare their performances. These three different wavelength assignment techniques are incorporated in the fitness function in term of the free wavelength factor $W_x$. While evaluating the fitness function if the wavelength continuity constraint is not satisfied for all the links of a chromosome, then $W_x$ is set to zero. If wavelength continuity constraint is satisfied for all the links of a chromosome, the free wavelength factor $W_x$ is made one.

**Wavelength constraints:**

First we define the following variables: $I_{ij}^{lp}$ as defined earlier tells whether a given link $(i,j)$ belongs to a lightpath; $I_{ijw}^{lp}$ is the lightpath wavelength indicator to show that the lightpath uses wavelength $w$ on link $(i,j)$; $I_{ijw}^{lp(x,y)}$ is the lightpath wavelength link indicator that is one when a lightpath uses wavelength $w$ on link $(i,j)$ between nodes $x$ and $y$; and $l^{(x,y)}$ is the physical link between nodes $x$ and $y$.

The wavelength constraints are as follows [14]:

$$I_{ij}^{lp} = \sum_{w=0}^{W-1} I_{ijw}^{lp} \ \forall (i,j) \qquad 2(a)$$

$$I_{ijw}^{lp(x,y)} \leq I_{ijw}^{lp} \ \forall (i,j),(x,y),w \qquad 2(b)$$

$$\sum_{i,j} I_{ijw}^{lp(x,y)} \leq 1 \ \forall (x,y),w \qquad 2(c)$$

$$\sum_{w=0}^{W-1}\sum_{x} I_{ijw}^{lp(x,y)} l^{(x,y)} - \sum_{w=0}^{W-1}\sum_{x} I_{ijw}^{lp(y,x)} l^{(y,x)} = I_{ij}^{lp}, y=j$$

$$\sum_{w=0}^{W-1}\sum_{x} I_{ijw}^{lp(x,y)} l^{(x,y)} - \sum_{w=0}^{W-1}\sum_{x} I_{ijw}^{lp(y,x)} l^{(y,x)} = -I_{ij}^{lp}, y=i \qquad 2(d)$$

$$\sum_{w=0}^{W-1}\sum_{x} I_{ijw}^{lp(x,y)} l^{(x,y)} - \sum_{w=0}^{W-1}\sum_{x} I_{ijw}^{lp(y,x)} l^{(y,x)} = 0, y \neq i, y \neq j$$

Equation 2(a) implies that the wavelength used by a lightpath is unique. Equation 2(b) ensures that the wavelength continuity constraint is adhered to. Equation 2(c) ensures that two lightpaths using the same link can not be assigned identical wavelengths. Equation 2(d) expresses the conservation of wavelengths at the end nodes of physical links on a lightpath.

## 3. THE PROPOSED RWA ALGORITHM

In this paper each chromosome, shown in Fig. 1, is a sequence of randomly generated nodes, called genes, satisfying the physical topology of the particular network. The chromosomes are supposed to have variable lengths. While generating the chromosomes, the topology database is utilized to ensure that each of the chromosomes represents a valid path from the source node to the destination node. In Fig.1, $k_1$ represents the neighbor of the source node $S$, $k_2$ represents the neighbor of $k_1$, and so on, till we get to $k_r$, which is a neighbor of the destination node *D*. Another point to be ensured while generating the chromosome is that the nodes are not repeated from any source to any destination; thus avoiding loops. While generating a chromosome a threshold time of $T_1$ is fixed for satisfying an initial hop-bound constraint for the chromosome. However, if no valid solution is found within this time limit, the bound is increased till a feasible solution is generated. This particular feature of the algorithm helps in reducing the overall execution time spent in the path finding process.

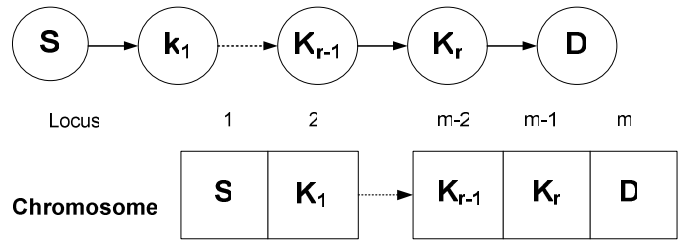

Fig.1. Chromosome representation

The main phases of the proposed EP algorithm are as follows: (i) Population initialization - the initial population size is considered to be one irrespective of the number of nodes in the network. This drastically reduces the execution time of the algorithm, which is one of our primary goals. The initial population or individual is generated randomly taking note of the fact that it represents a valid path from the source node to the destination node and that the problem constraints are satisfied; (ii) Mutation - the mutation of chromosomes takes place by changing or flipping one of the genes of the candidate chromosome. In the proposed algorithm mutation probability is kept equal to one. Mutation site of the parent chromosome is chosen randomly and from that site to the destination node a partial new path is generated based on the topological database. While implementing the algorithm, it is seen that the nodes do not get repeated, avoiding undesirable loops in the path. Care should also be taken such that the offspring chromosomes satisfy the initial hop-count bound constraint for a threshold time of T2 seconds. If a feasible path is not found within that time the hop bound is relaxed for that request. Before mutation the encoding of a chromosome is as shown in Fig.2. Fig. 3 shows the newly generated chromosome after mutation if the mutation site comes out to be the position three or locus three where node $k_4$ is located. The encoding of the chromosome remains same till locus three or node $k_4$ and from $k_4$ onwards the nodes are randomly generated based on the topology connectivity matrix till we reach the destination node D. This way a single parent undergoes the mutation loop fifteen times to produces fifteen offsprings. The chromosome with the highest fitness value survives for the next generation. (iii) Selection – in this step the chromosome with the highest fitness value is selected and get copied into the next generation. Step (ii) and step (iii) are repeatedly executed till certain termination criterion is reached.

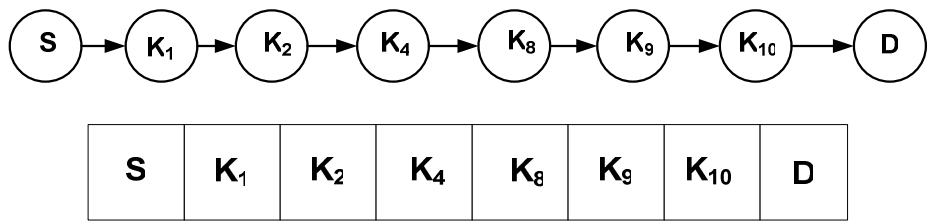

Fig.2. Chromosome representation before mutation

International journal of Computer Networks & Communications (IJCNC), Vol.2, No.2, March 2010

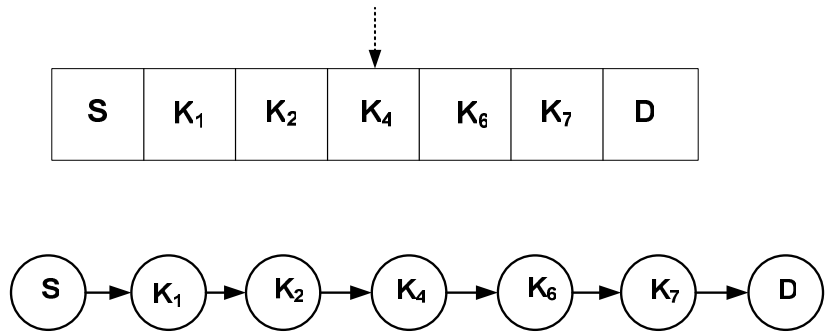

Fig.3. Chromosome representation after mutation

**Wavelength Assignment**

The flow chart for the resulting RWA algorithm is shown in Fig. 4.

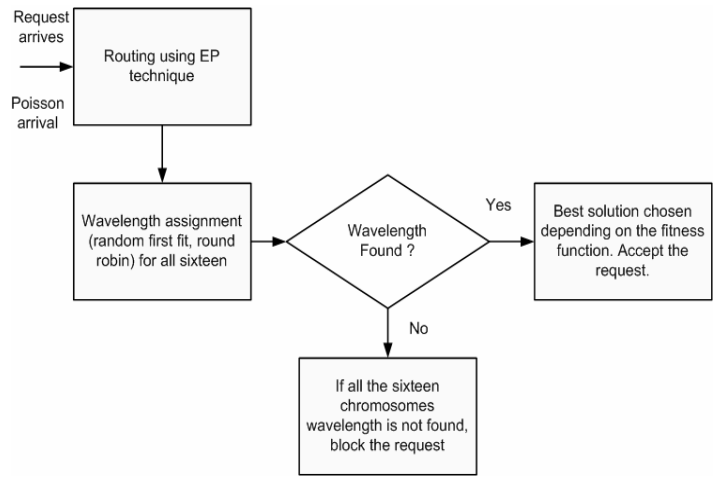

Fig.4. Flow chart of routing and wavelength assignment algorithm

### 4. THE ALGORITHM IMPLEMENTATION

The 14-node NSF network topology [9], is used to evaluate the algorithm. The cost values are assigned as in the reference [9]. The parameters $T_1$, $T_2$, are kept at 0.5sec, 1.5sec, respectively. These values were determined experimentally by trial and error to get an overall low mean execution time.

In our simulation the source destination pair is randomly chosen according to a Uniform distribution. We have calculated the upper bound for the mean execution time as suggested by Bisbal et al. in [9]. Within the threshold times of $T_1$ and $T_2$, the proposed algorithm tries to find a feasible route that satisfies the hop bound constraint, which is kept at four in our experiment. Once the lightpath is set up, the set up time of the light path is calculated and is incorporated in the fitness function. In our algorithm we have a centralized controller to update the database of available wavelengths for each of the links. Since evolutionary programming is a randomized algorithm, to correctly assess the result we have executed the program in each of the simulation experiments ten times.

## 5. SIMULATION RESULTS AND DISCUSSIONS:

The proposed technique was simulated using Microsoft Visual C++ on an Intel Centrino processor (1.6 GHz clock and 512 MB RAM, 20 GB HDD). Approximately $10^6$ to $10^7$ requests are generated during the simulation. The mean blocking probabilities obtained by the proposed algorithm for three different wavelength assignment techniques are shown in Fig.5 for a traffic load of 60 Erlang and a total of eight wavelengths per link. At seventh and eighth generation the mean blocking probability converges for all the three types of wavelength assignment techniques and is superior compared to [9, 10, 11]. Out of the three types of wavelength assignment techniques, round robin shows the best performance for all the generations for different types of wavelength assignment techniques.

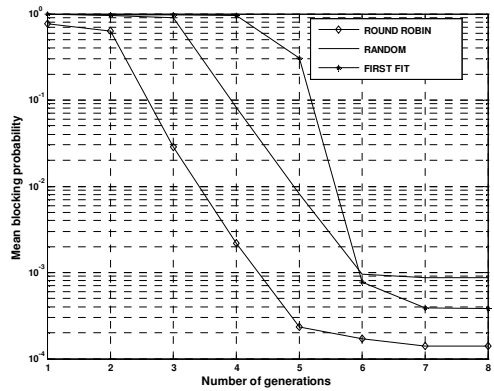

Fig.5 Mean blocking probability of EP with W=8 and a traffic load of 60 Erlang.

Fig.6. shows the mean execution time obtained by the proposed algorithm for three different wavelength assignment techniques. Among all the three wavelength assignment techniques, first fit exhibits the lowest mean execution time compared to the other techniques. Round robin exhibits the worst performance in terms of the execution time.

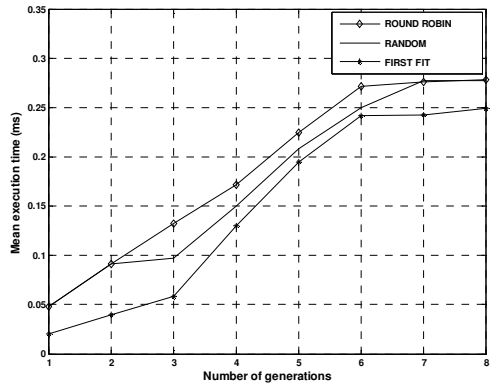

Fig.6 Mean execution time (ms) of EP with W=8 and a traffic load of 60 Erlang.

Fig.7. shows the mean blocking probability for all the three wavelength assignment techniques when the holding time is Pareto distributed. The plot is observed for a network load of 60 Erlang. From the plot, it is observed that all the three wavelength assignment approaches exhibit slightly less mean blocking probability compared to the values obtained when the holding time is exponentially distributed. For our experiment we have kept the shape parameter value equal to 1.2 and location parameter value equal to 1. Among the three wavelength assignment techniques, round robin performs the best in terms of the mean blocking probability and random assignment technique shows the worst performance. From this study it is concluded that the mean blocking probability is insensitive to the type of distribution of the holding time.

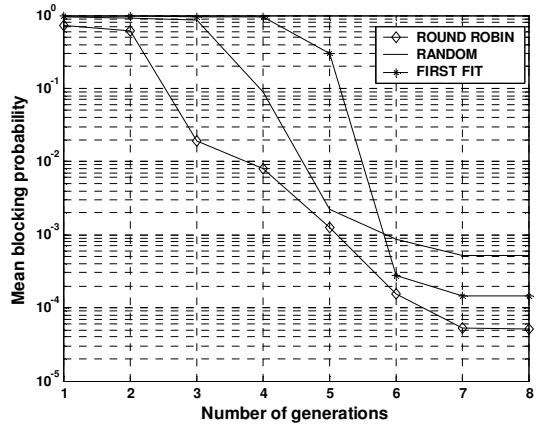

Fig.7 Mean blocking probability of EP for Pareto holding time with W=8 and traffic load of 60 Erlang.

Fig.8. shows the comparison of the mean execution time for three different wavelength assignment techniques when the holding time is Pareto distributed. The execution time shown for three wavelength assignment techniques are more or less similar to the execution time obtained when the simulation is carried for exponential holding time. However, the mean execution time obtained by first fit assignment technique is smaller compared to the other two techniques.

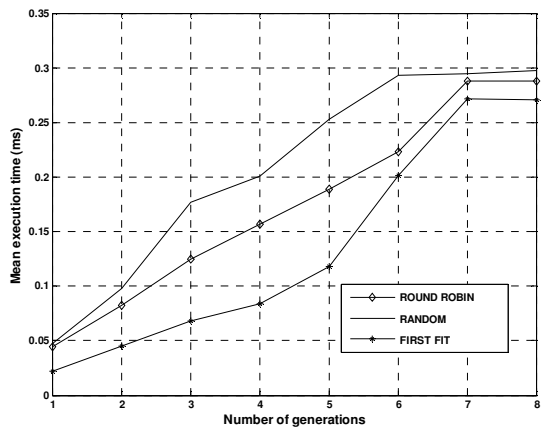

Fig.8 Mean execution time of EP for the Pareto holding time with W=8 and traffic load of 60 Erlang.

In Fig.9, we show the comparison of mean blocking probability between the exponential holding time and the Pareto holding time for the round robin wavelength assignment technique. The mean of arrival rate is kept same and the mean of holding time of each process is also kept same for a fair comparison of both the processes. From the plot it is observed that the mean blocking probability for the Pareto process is little better compared to the exponential holding time for all the traffic loads. The difference in the mean blocking probability is more when the network traffic load is less than or equal to 80 Erlang.

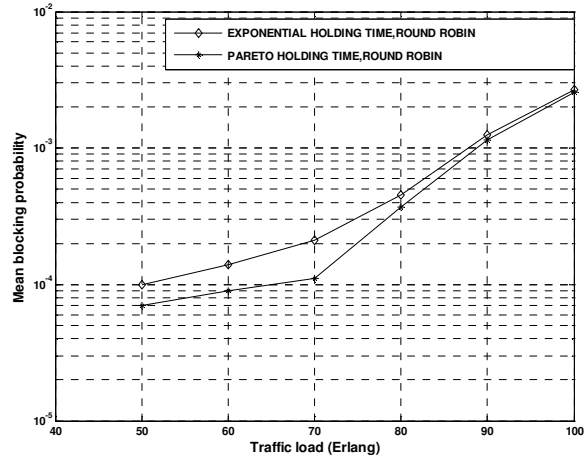

Fig.9. Mean blocking probability of EP for the exponential and Pareto holding times for the DRWA problem with W=8 at various traffic loads.(Round robin wavelength assignment)

Fig.10. shows that the mean execution time exhibited by the proposed algorithm for each of the exponential and Pareto holding times. We observe that a the curve for a Pareto exhibits slightly higher mean execution time compared to exponential for all the traffic loads. This experiment is done only for the first fit wavelength assignment technique. It is found from our previous experiment, that this technique exhibits lesser execution time compared to round robin and random assignment techniques. In this experiment we have fixed the value of location parameter to 1.5 and the shape parameter value is changed according to the load.

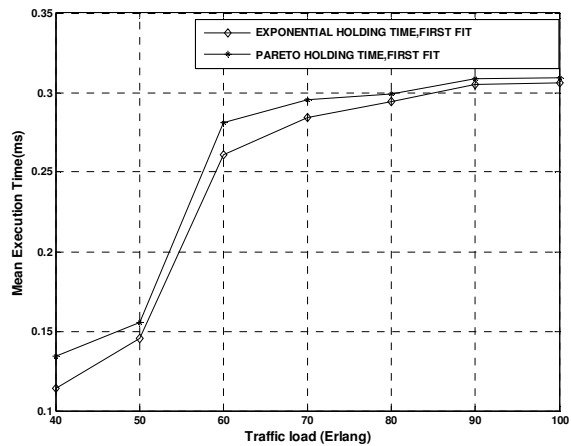

Fig.10. Mean execution time of EP for both exponential and Pareto holding time for the DRWA problem with W=8 and various traffic loads. (first fit wavelength assignment).

International journal of Computer Networks & Communications (IJCNC), Vol.2, No.2, March 2010

We have also estimated the number of wavelengths required by the DRWA algorithm to obtain the minimum average blocking probability for both exponentially and Pareto distributed holding time. From the figure11.and 12. we observe that 12 wavelengths are sufficient to obtain fairly low mean blocking probability.

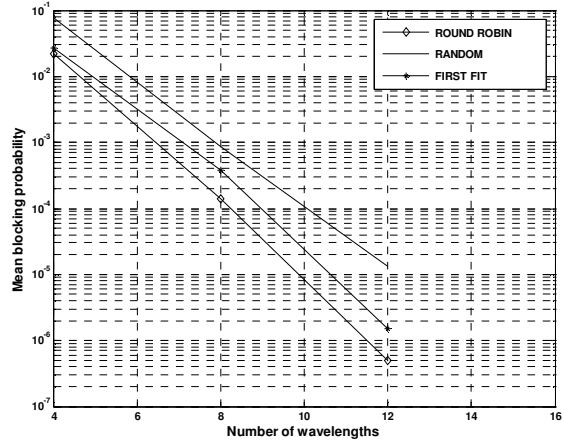

Fig.11. Mean blocking probability of EP for the exponential holding time for the DRWA problem with various wavelengths.

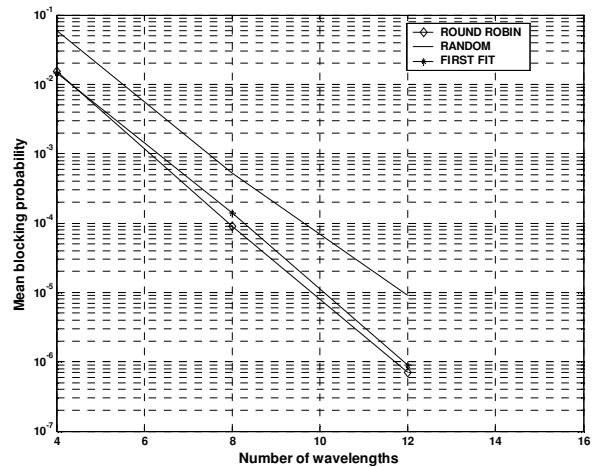

Fig.12. Mean blocking probability of EP for the Pareto holding time for the DRWA problem with various wavelengths.

**Algorithm time complexity analysis for the DRWA problem**

Let us assume that $N$ denotes the number of nodes in the network, $P$ and $G$ denote the population size and the number of iterations respectively, and $C$ denotes the number of offsprings. Let W be the number of wavelengths per link.

We compute the overall worst case time complexity of the algorithm using a methodology similar to [9].

International journal of Computer Networks & Communications (IJCNC), Vol.2, No.2, March 2010

*Step 1:* The time taken to generate one random route and to evaluate its cost together will be equal to $O(N)+O(N) = O(N)$ time units.

*Step 2:* The complexity of generation of a random route (chromosome) and its offsprings is $O(C \cdot N)$ time units. Then cost of evaluation or fitness evaluation takes $O(W \cdot C \cdot N)$ time units. In the worst case, all wavelengths in all the links of the route (this number is limited to a maximum of N-1 links) are examined. So total complexity in this step is equal to $O(C \cdot N) + O(W \cdot C \cdot N) = O(W \cdot C \cdot N)$ time units.

*Step 3:* In this step the best individual is sorted out of *P* chromosomes, where *P=(C+1)*. This involves sorting the population in decreasing order of fitness values; so, the complexity of this operation is *O(PlogP)*.

Overall complexity: $O(N) + O(G \cdot ((W \cdot C \cdot N) + PlogP)) = O(G \cdot W \cdot C \cdot N)$ time units. This is less than the overall complexity of the approach reported in [9] and [11].

**Number of fitness function evaluation**

In our algorithm the number of fitness function evaluation is equal to $1+(G \cdot C) \approx (G \cdot C)$.

## 6. CONCLUSION:

In this paper we have proposed a novel solution to the DRWA problem based on an evolutionary programming algorithm. The proposed solution is evaluated with different wavelength assignment techniques. We conclude that the proposed solution incorporating a round robin wavelength assignment technique can be used to enhance the operation of current WDM lightwave networks.

**Authors**

**Urmila Bhanja** is presently, a PhD candidate in the department of Electrical & Electronics Communication Engineering, Indian Institute of Technology, Kharagpur, West Midnapur, West Bengal, India. Her research interest includes computer network design resource optimization using bio-inspired algorithms, development of different soft computing approaches for function optimization etc.

**Dr.Sudipta Mahapatra** received the B.Sc. (Engg.) degree in Electronics and Telecommunications Engineering from Sambalpur University, India, in 1990, and the M.Tech. and Ph.D. degrees in Computer Engineering from Indian Institute of Technology (IIT), Kharagpur, India, in 1992 and 1997, respectively. From April 1993 to September 2002, he worked in various capacities in the Computer Science and Engineering Department, of National Institute of Technology (NIT), Rourkela, India. He visited the Electronic Systems Design Group of Loughborough University, U.K., as a BOYSCAST Fellow of the Department of Science and Technology, Government of India, from March 1999 to March 2000. Currently he is working as an Associate Professor in the Department of Electronics and Electrical Communication Engineering, IIT, Kharagpur, India. His areas of research interest include parallel and distributed computing, computer networking and data compression hardware.

**Dr.Rajarshi Roy** did his PhD in Electrical Engineering from Polytechnic University, Brooklyn, NY, USA in 2001. He is an assistant professor in Dept. of E & ECE, Indian Institute of Technology, Kharagpur, West Midnapur, West Bengal, India. He did his M.Sc. (Engg.) from Indian Institute of Science, Bangalore, Karnataka, India in 1995 and Bachelors in Electronics and telecommunication Engg. from Jadavpur University, Kolkata, West Bengal, India in 1992.  He did work for Comverse, USA and Lucent, India. He served Helsinki University of Technology, Finland and Indian Statistical Institute, Kolkata, India as academic visitor and Bell Labs, Holmdel, NJ, USA as a summer intern. His area of interest includes Communication networks, Social networking, Operation research and Distributed computing.